\documentclass[twocolumn,showpacs,preprintnumbers,amsmath,amssymb]{revtex4}

\usepackage{graphicx}% Include figure files
\usepackage{dcolumn}% Align table columns on decimal point
\usepackage{bm}% bold math

\begin{document}
\preprint{APS/123-QED}

\title{SCATTERING OF COSMIC RAYS BY MAGNETOHYDRODYNAMIC INTERSTELLAR 
TURBULENCE}
\author{Huirong Yan}
%\altaffiliation[Also at ]{Physics Department, XYZ University.}%Lines break automatically or can be forced with \\

\author{A. Lazarian}%
 %\email{Second.Author@institution.edu}
\affiliation{%
Astronomy Dept., Univ.~of Wisconsin, Madison, WI53706, USA; %\\
 yan, lazarian@astro.wisc.edu
%This line break forced with \textbackslash\textbackslash
}%

\date{\today}% It is always \today, today,
             %  but any date may be explicitly specified

\begin{abstract}
Recent advances in understanding of magnetohydrodynamic (MHD) turbulence
call for substantial revisions in our understanding of cosmic ray
transport. In this paper we use gyroresonance recently obtained scaling laws for MHD modes to calculate the scattering frequency for cosmic rays in the ISM. We consider 
gyroresonance with MHD modes (Alfv\'{e}nic, slow and fast) and transit-time
damping (TTD) by fast modes. We conclude that the gyroresonance with fast modes
is the dominant contribution to cosmic ray scattering for the typical interstellar conditions. In contrast to earlier studies, we find that Alfv\'{e}nic and
slow modes are inefficient because they are far from isotropy usually assumed. 
\end{abstract}

% PACS, the Physics and Astronomy
                             % Classification Scheme.
%\keywords{Suggested keywords}%Use showkeys class option if keyword
                              %display desired
\maketitle

\section{\label{sec:intro}INTRODUCTION }

The propagation of cosmic rays (CRs) is affected by their interaction
with magnetic field. This field is turbulent and therefore, the resonant interaction
of cosmic rays with MHD turbulence has been discussed by many authors
as the principal mechanism to scatter and isotropize cosmic rays (\cite{Sch02}). Although cosmic ray diffusion can happen while cosmic rays follow wandering magnetic fields \cite{Jok}, the acceleration of cosmic rays requires efficient scattering.

While most investigations are restricted to Alfv\'{e}n waves propagating
along an external magnetic field (the so-called slab model
of Alfv\'{e}nic turbulence), obliquely propagating MHD waves have been
included in \cite{Fisk} and later studies \cite{Bieber,PP}. The problem, however, is that the Alfv\'{e}nic turbulence considered in their studies is isotropic turbulence, and this is contrary to the modern
understanding of MHD turbulence (\cite{GS95}, see \cite{CLY} for a review and references therein).

A recent study \cite{Lerche} found a strong dependence of scattering on turbulence anisotropy.
 Therefore the calculations of CR scattering must be done using a realistic MHD turbulence model.
  An important attempt in this direction was carried out in \cite{Chana}. 
  There Alfv\'{e}n modes were
treated 
in the spirit of Goldreich-Shridhar \cite{GS95} (1995, henceforth GS95) model 
  of incompressible turbulence and marginal scattering was obtained. 
However, a more accurate description is now available (\cite{CLV}) and 
thus there is a need to revisit the problem. Moreover,
\cite{Chana} did not consider compressible modes, while we show below that 
these modes provide the dominant contribution to the scattering.

\section{MHD STATISTICS}

MHD perturbations can be decomposed into Alfv\'{e}nic, slow and fast modes (see \cite{Ginz}). Alfv\'{e}nic turbulence is considered by many authors as the
default model of interstellar magnetic turbulence. This is partially
motivated by the fact that unlike compressible modes, the Alfv\'{e}n ones are
essentially free of damping in fully ionized medium (see \cite{Ginz,Kulsrud}).

Unlike hydrodynamic turbulence, Alfv\'{e}nic turbulence is anisotropic,
with eddies elongated along the magnetic field. This happens because
it is easier to mix the magnetic field lines perpendicular to the
direction of the magnetic field rather than to bend them. 
The GS95 model describes {\it incompressible} Alfv\'{e}nic turbulence, which
formally means that plasma \( \beta =P_{gas}/P_{mag}=2C_{s}^{2}/V_{A}^{2} \)
is infinity. It was first conjectured in \cite{LG01} that
GS95 scaling should be approximately true for moderately compressible plasma. For low \( \beta  \) plasma Cho \& Lazarian \cite{CL}
(henceforth CL02) showed that
the coupling of Alfv\'{e}nic and compressible modes is weak and that
the Alfv\'{e}nic modes follow the GS95 spectrum \cite{coup}. This is consistent with
the analysis of observational data (\cite{LP,SL}). In what follows, we consider both Alfv\'{e}n modes and compressible modes
and use the description of those modes obtained in CL02 to study CR scattering
by MHD turbulence in a medium with energy injection scale \( L=100 \)pc,
density \( n=10^{-4} \)cm\( ^{-3}, \) temperature \( T=2\times 10^6 \)K. Recent observations  \cite{Beck} suggest that matter in the galactic halos
is magnetic-dominant, corresponding to low \(\beta\) medium, here we choose \( \beta \simeq 0.1 \). The injection length scale is important as 
Alfv\'{e}nic turbulence 
exhibits scale-dependent anisotropy that increases with the decrease of the scale.

We describe MHD turbulence statistics by correlation functions. Using
the notations from \cite{Chana}, we get the expressions for the
correlation tensors in Fourier space

\begin{eqnarray}
<B_i(\mathbf{k})B_j^{*}(\mathbf{k'})>/B_{0}^{2}=\delta (\mathbf{k}-\mathbf{k'})M_{ij }(\mathbf{k}), \nonumber \\
<v_i(\mathbf{k})B_j^{*}(\mathbf{k'})>/V_{A}B_{0}=\delta (\mathbf{k}-\mathbf{k'})C_{ij }(\mathbf{k}), \nonumber \\
<v_i(\mathbf{k})v_j^{*}(\mathbf{k'})>/V_{A}^{2}=\delta (\mathbf{k}-\mathbf{k'})K_{ij }(\mathbf{k}),
\end{eqnarray}
 where \( B_{\alpha,\beta} \) is the magnetic field fluctuations.

The isotropic tensor usually used in the literature is
\begin{equation}
\label{isotropic}
K_{ij}(\mathbf{k})=C_0\{\delta_{ij}-k_ik_j/k^2\}k^{-11/3},
\end{equation}
The normalization constant \(C_0\) can be obtained if the energy input at the scale \(L \) is defined. 
Assuming equipartition, the kinetic energy density \(\epsilon_k=\int dk^3 \sum_{i=1}^3 K_{ii} \rho V_A^2/2 \sim B_0^2/8\pi \), we get \( C_0=L^{-2/3}/12\pi \). 
The analytical fit to the anisotropic tensor for Alfv\'{e}n modes, obtained in \cite{CLV} is,
\begin{equation}
\label{anisotropic}
K_{ij}(\mathbf{k})=C_aI_{ij}k_{\perp }^{-10/3}exp(-L^{1/3}k_{\parallel }/k_{\perp }^{2/3}),
\end{equation}
where \( I_{ij}=\{\delta_{ij}-k_ik_j/k_{\perp }  ^2\} \) is a 2D matrix in x-y plane, \( k_{\parallel } \) is the wave vector along the local mean
magnetic field (see\cite{CLY}), \( k_{\perp } \) is the wave
vector perpendicular to the magnetic field and the normalization constant \( C_a=L^{-1/3}/6\pi \). 
The tensors in \cite{Chana} used step function instead of the exponent.  We assume that for the Alfv\'{e}n modes
\( M_{ij }=K_{ij }, \) \( C_{ij }=\sigma M_{ij } \)
where the fractional helicity \( -1<\sigma <1 \) is independent of
\( \mathbf{k} \) (\cite{Chana}).

Numerical calculations in CL02 demonstrated that slow modes follow GS95
scalings. The correlation tensors for slow modes in low \(\beta\) plasma are \cite{tensor}
\begin{equation}
\left[ \begin{array}{c}
M_{ij}({\mathbf{k}})\\
C_{ij}({\mathbf{k}})\\
K_{ij}({\mathbf{k}})
\end{array}\right] ={C_a\beta^2\over 16}  \sin^2(2\theta) J_{ij}k_\perp^{-{10\over3}}exp(-{L^{{1\over 3}}k_{\parallel }\over k_{\perp }^{{2\over3}}})\left[ \begin{array}{c}
\cos^2\theta \\
\sigma\cos\theta \\
1
\end{array}\right]\nonumber,
\end{equation}
where \(\cos\theta=k_\parallel/k \),\( J_{ij}=k_{i}k_{j}/k_{\perp }^{2} \) is also a 2D tensor in \(x-y\) plane .

According to CL02, fast modes are isotropic and have one dimensional spectrum \( E(k)\propto k^{-3/2} \). 
In low \( \beta \) medium, the velocity fluctuations are always perpendicular to \( \mathbf{B}_{0} \) for all \( \mathbf{k} \), 
while the magnetic fluctuations are perpendicular to \( \mathbf{k} \). Thus \( K_{ij }, M_{ij } \) of fast modes are not equal, their x-y components are \cite{diffSch}
\begin{equation}
\left[ \begin{array}{c}
M_{ij}({\mathbf{k}})\\
C_{ij}({\mathbf{k}})\\
K_{ij}({\mathbf{k}})
\end{array}\right] ={L^{-1/2}\over 8\pi} J_{ij}k^{-7/2}\left[ \begin{array}{c}
\cos^{2}\theta \\
\sigma \cos\theta \\
1
\end{array}\right] ,
\end{equation}

 In high \( \beta \) medium, the velocity fluctuations are radial, i.e., 
along the direction of \( {\bf k} \). Fast modes in this regime are 
essentially 
sound waves compressing magnetic field (\cite{GS95}, \cite{LG01}, Cho \& Lazarin, in preparation). 
The compression of magnetic field depends on plasma \(\beta \). The 
corresponding x-y components of the tensors are
\begin{equation}
\left[ \begin{array}{c}
M_{ij}({\mathbf{k}})\\
C_{ij}({\mathbf{k}})\\
K_{ij}({\mathbf{k}})
\end{array}\right] ={L^{-1/2}\over 8\pi} \sin^2\theta J_{ij}k^{-7/2}\left[ \begin{array}{c}
\cos^2\theta /\beta\\
\sigma \cos\theta/\beta^{1/2} \\
1
\end{array}\right] .
\end{equation}

\section{SCATTERING BY ALFV\'{E}NIC TURBULENCE}

Particles get into resonance with MHD perturbations propagating along the
magnetic field if the resonant condition is fulfilled, namely, \( \omega =k_{\parallel }v\mu +n\Omega  \),(\( n=\pm 1,2... \))
where \( \omega  \) is the wave frequency, \( \Omega =\Omega _{0}/\gamma  \)
is the gyrofrequency of relativistic particle, \( \mu =\cos \alpha  \), where \( \alpha \) is the pitch angle of particles. In other words, resonant interaction between a particle and the
transverse electric field of a wave occurs when the Doppler shifted
frequency of the wave in the particle's guiding center rest frame
\( \omega _{gc}=\omega -k_{\parallel }v\mu  \) is a multiple of the
particle gyrofrequency. For high energy particles, the resonance happens for
both positive and negative \( n \).

We employ quasi-linear theory (QLT) to obtain our estimates. QLT has been proved to be a useful tool in spite of its intrinsic limitations (\cite{Chana, Sch98, Mi97}). For moderate energy cosmic rays, the corresponding resonant scales are much smaller than the injection scale. 
Therefore the fluctuation on the resonant scale \( \delta B\ll B_0 \) even if they are comparable at the injection scale. QLT disregards diffusion of cosmic rays that follow wandering magnetic field lines (\cite{Jok}) and this diffusion should be accounted separately.
  Obtained by applying the QLT to the collisionless Boltzmann-Vlasov
equation, the Fokker-Planck equation is generally used to describe
the involvement of the gyrophase-average distribution function \( f \),

\[
\frac{\partial f}{\partial t}=\frac{\partial }{\partial \mu }\left( D_{\mu \mu }\frac{\partial f}{\partial \mu }+D_{\mu p}\frac{\partial f}{\partial p}\right) +\frac{1}{p^{2}}\frac{\partial }{\partial p}\left[ p^{2}\left( D_{\mu p}\frac{\partial f}{\partial \mu }+D_{pp}\frac{\partial f}{\partial p}\right) \right] ,\]
where \( p \) is the particle momentum. The Fokker-Planck coefficients
\( D_{\mu \mu },D_{\mu p},D_{pp} \) are the fundamental physical
parameter for measuring the stochastic interactions, which are determined
by the electromagnetic fluctuations\cite{Sch93}:

\begin{eqnarray}
<B_i(\mathbf{k})B_j^{*}(\mathbf{k'})>=\delta (\mathbf{k}-\mathbf{k'})P_{ij }(\mathbf{k}), \nonumber \\
<B_i(\mathbf{k})E_j^{*}(\mathbf{k'})>=\delta (\mathbf{k}-\mathbf{k'})T_{ij }(\mathbf{k}), \nonumber \\
<E_i(\mathbf{k})B_j^{*}(\mathbf{k'})>=\delta (\mathbf{k}-\mathbf{k'})Q_{ij }(\mathbf{k}), \nonumber \\
<E_i(\mathbf{k})E_j^{*}(\mathbf{k'})>=\delta (\mathbf{k}-\mathbf{k'})R_{ij }(\mathbf{k}).
\end{eqnarray}

 From Ohm's Law \( \mathbf{E}(\mathbf{k})=-(1/c)\mathbf{v}(\mathbf{k})\times \mathbf{B}_{0}, \)
we can express the electromagnetic fluctuations \( T_{ij },R_{ij } \)
in terms of correlation tensors \( C_{ij }, \) \( K_{ij } \). Adopting the approach in \cite{Sch93}, we can get the Fokker-Planck coefficients in the lowest order approximation of \( V_{A}/c \),

\begin{eqnarray}
\left[ \begin{array}{c}
D_{\mu \mu }\\
D_{\mu p}\\
D_{pp}
\end{array}\right] = {\Omega ^{2}(1-\mu ^{2})\over 2B_{0}^{2}}\left[ \begin{array}{c}
1\\
mc\\
m^{2}c^{2}
\end{array}\right] {\mathcal{R}}e \sum _{n=-\infty }^{n=\infty }\int_{k_{min}}^{k_{max} } dk^3  \nonumber \\
\int _{0}^{\infty }dte^{-i(k_{\parallel }v_{\parallel }-\omega +n\Omega )t} \left\{ J_{n+1}^{2}({k_{\perp }v_{\perp }\over \Omega })\left[ \begin{array}{c}
P_{{\cal RR}}({\mathbf{k}})\\
T_{{\cal RR}}({\mathbf{k}})\\
R_{{\cal RR}}({\mathbf{k}})
\end{array}\right] \right. \nonumber \\
+ J_{n-1}^{2}({k_{\perp }v_{\perp }\over \Omega }) \left[ \begin{array}{c}
P_{{\cal LL}}({\mathbf{k}})\\
-T_{{\cal LL}}({\mathbf{k}})\\
R_{{\cal LL}}({\mathbf{k}})
\end{array} \right] + J_{n+1}({k_{\perp }v_{\perp }\over \Omega })J_{n-1}({k_{\perp }v_{\perp }\over \Omega })\nonumber \\
\left. \left[ e^{i2\phi }\left[ \begin{array}{c}
-P_{{\cal RL}}({\mathbf{k}})\\
T_{{\cal RL}}({\mathbf{k}})\\
R_{{\cal RL}}({\mathbf{k}})
\end{array}\right] +e^{-i2\phi }\left[ \begin{array}{c}
-P_{{\cal LR}}({\mathbf{k}})\\
-T_{{\cal LR}}({\mathbf{k}})\\
R_{{\cal LR}}({\mathbf{k}})
\end{array}\right] \right] \right\} \label{genmu} 
\end{eqnarray}
where  \( k_{min}=L^{-1} \), \( k_{max}=\Omega _{0}/v_{th} \) corresponds to the dissipation scale, \(m=\gamma m_H\) is the relativistic mass of the proton, \( v_{\perp } \) is the particle's velocity component perpendicular
to \( \mathbf{B}_{0} \), \( \phi =\arctan (k_{y}/k_{x}), \) \( {\cal L},{\cal R}=(x\pm iy)/\sqrt{2} \)
represent left and right hand polarization\cite{Dup}.

The integration over time gives us a delta function \( \delta (k_{\parallel }v_{\parallel }-\omega +n\Omega ) \),
corresponding to  static magnetic perturbations (\cite{Sch93,static}). 
For cosmic rays, \( k_{\parallel }v_{\parallel }\gg \omega =k_{\parallel }V_{A} \)
so that the resonant condition is just \( k_{\parallel }v\mu +n\Omega =0 \).
From this resonance condition, we know that the most important interaction
occurs at \( k_{\parallel }=k_{res}=\Omega /v_{\parallel } \).

Noticing that the integrand
for small \( k_{\perp } \) is substantially suppressed by the exponent
in the anisotropic tensor (see Eq. (\ref{anisotropic})) so that the
large scale contribution is not important, we can simply use the asymptotic
form of Bessel function for large argument. Then if the pitch 
angle \( \alpha \) not close to 0,
we can derive the analytical result for anisotropic turbulence,
\begin{equation}
\label{ana}
\left[ \begin{array}{c}
D_{\mu \mu }\\
D_{\mu p}\\
D_{pp}
\end{array}\right] =\frac{v^{2.5}\cos \alpha ^{5.5}}{2\Omega^{1.5} L^{2.5}\sin\alpha}\Gamma [6.5,k_{max}^{-\frac{2}{3}}k_{res}L^{\frac{1}{3}}]\left[ \begin{array}{c}
1\\
\sigma mV_{A}\\
m^{2}V_{A}^{2}
\end{array}\right] ,
\end{equation}
 where \( \Gamma [a,z] \) is the incomplete
gamma function. The presence of this gamma function in our solution makes our results orders of magnitude larger than those in \cite{Chana} for the most of energies considered (see Fig.1a). However, the scattering frequency \( \nu =2D_{\mu \mu }/(1-\mu ^{2}) \) are much smaller than the estimates for isotropic model. Unless we consider very high energy CRs (\( \ge 10^8 GeV \)) with the corresponding Larmor radius comparable to the turbulence injection scale, we can neglect scattering by the Alfv\'{e}nic turbulence.
What is the alternative way to scatter cosmic rays?

\begin{figure}[t]
\includegraphics[width=.49\columnwidth]{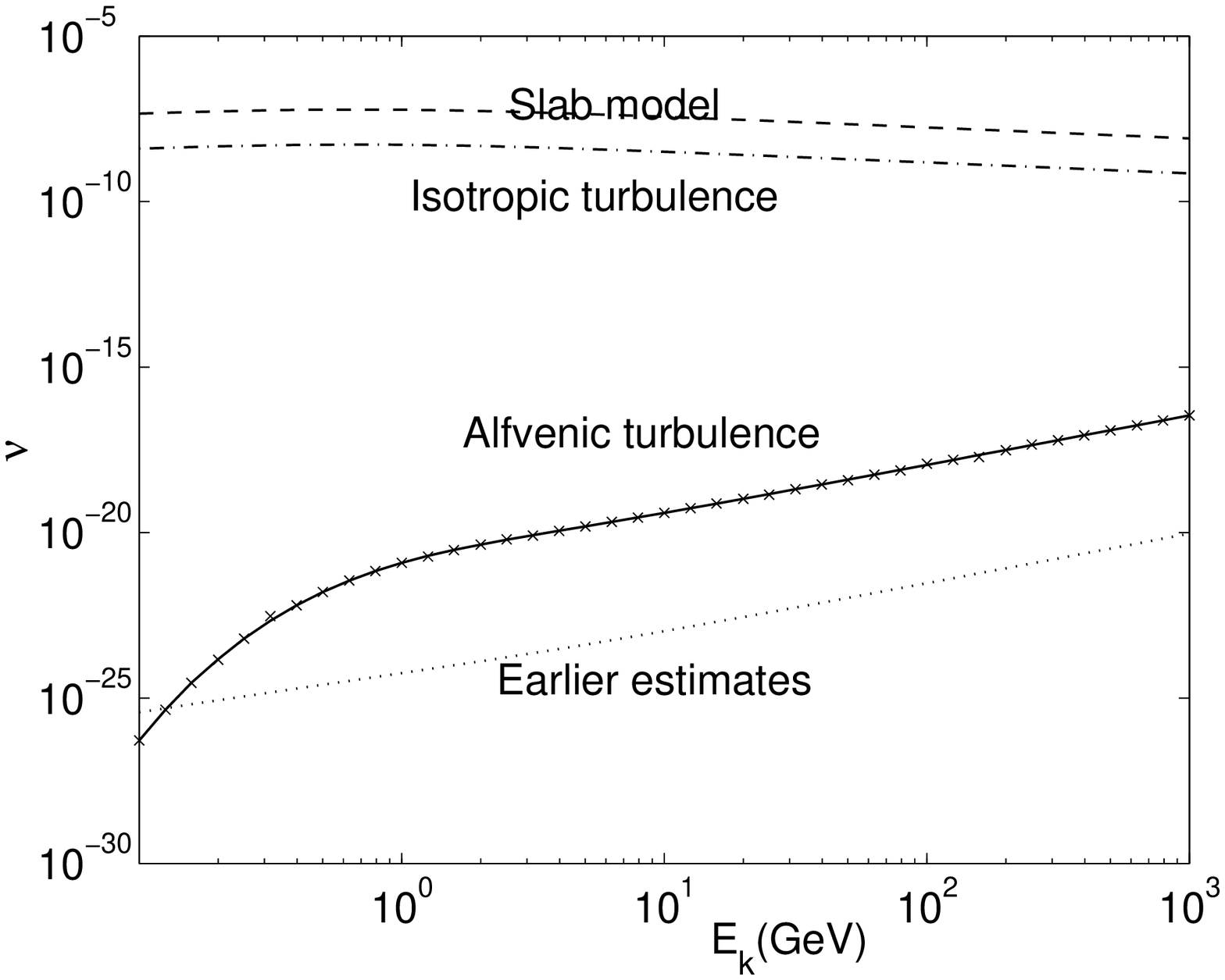} \hfil
\includegraphics[width=.49\columnwidth]{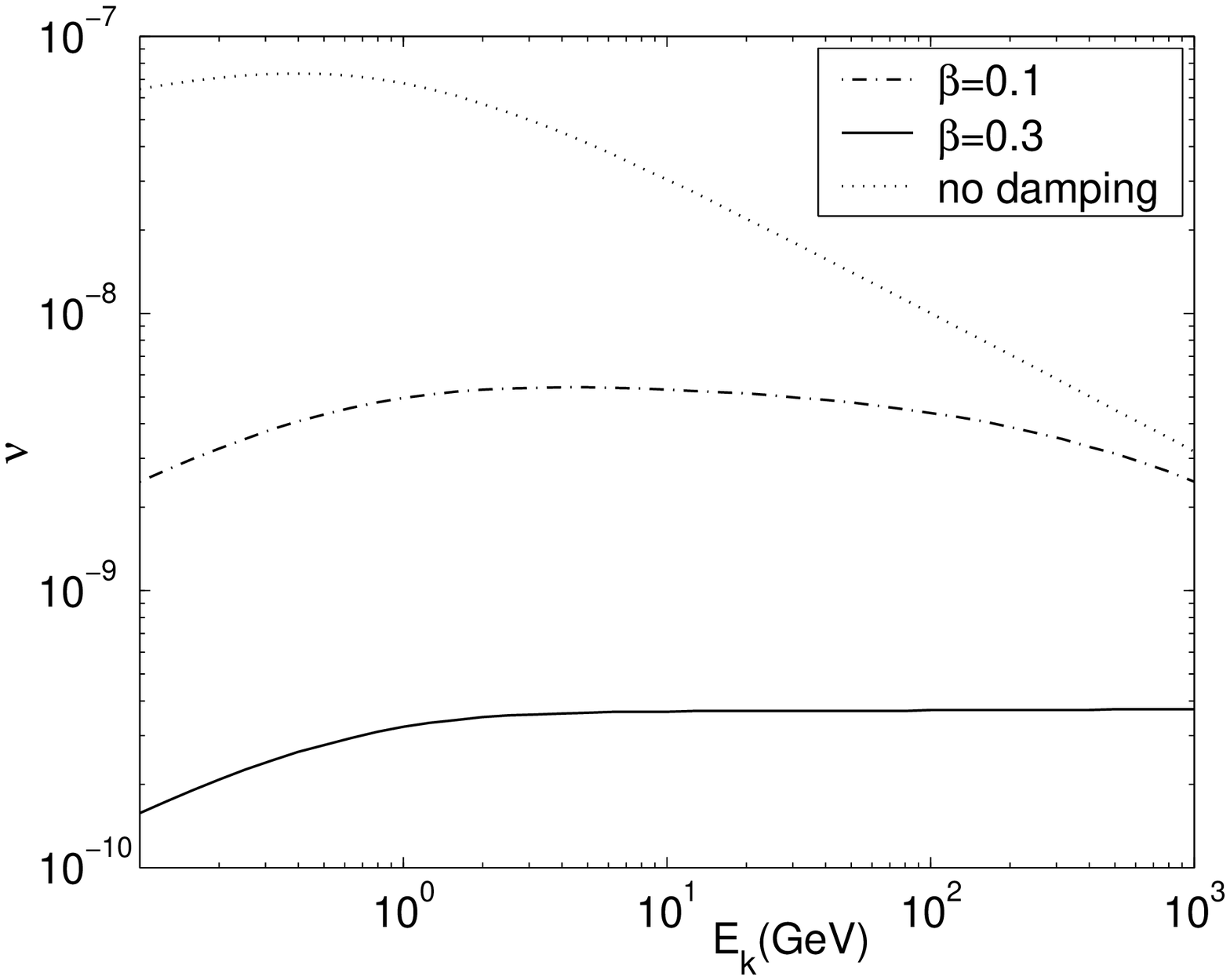}
\caption{The scattering frequency  \(\nu \) vs. the kinetic energy \( E_k \) of cosmic rays (a) by Alfv\'{e}nic turbulence,
(b) by fast modes. In (a), the dash-dot line refers to the scattering
frequency for isotropic turbulence. The '\protect\protect\protect\protect\( \times \protect \protect \protect \protect \)'
represents our numerical result for anisotropic turbulence, the solid
line is our analytical result from Eq.(\ref{ana}). Also plotted (dashed
line) is the previous result for anisotropic turbulence in \cite{Chana}. In (b), the dashed line represents the scattering by fast modes immune from damping, the solid and dashdot line are the results taking into account collisionless damping. 
}
\label{fig:incom}
\end{figure}

\section{SCATTERING by fast modes}

Our result that anisotropic turbulence is inefficient in CR scattering agrees well with the conclusions reached in \cite{Chana} and \cite{Lerche}. The contribution from slow modes is not larger than that by Alfv\'{e}n modes since the slow modes have the similar anisotropies and scalings. More promising are fast modes, which are isotropic (\cite{CL}). For fast modes we discuss two types of resonant interaction: gyroresonance and transit-time damping; the latter requires longitudinal motions. However, fast modes are subject to collisionless
damping which suppresses scattering\cite{damp}. The damping rate \(\gamma _{d}=\tau _{d}^{-1}\) for the
low \( \beta  \) case (\cite{Ginz}) is

\begin{eqnarray}
\label{lan}
\gamma _{d} &=& \frac{\sqrt{\pi\beta} }{4}V_{A}k\frac{\sin ^{2}\theta }{\cos \theta }\times [\sqrt{\frac{m_e}{m_H}}\exp(-\frac{m_e}{m_H\beta\cos ^{2}\theta }) \nonumber \\
&+& 5\exp(-\frac{1}{\beta\cos ^{2}\theta })],
\end{eqnarray}
 where \( m_{e} \) is the
electron mass. We see that the damping increases with \( \beta \). According to CL02, fast modes cascade over time
scales \( \tau _{fk}=\tau _{k}\times V_{A}/v_{k}=(k\times k_{min})^{-1/2}\times V_{A}/V^{2}, \)
where \( \tau _{k}=(kv_{k})^{-1} \) is the eddy turn-over time, \( V \) is the turbulence velocity at the injection scale.

Consider gyroresonance scattering in the presence of collisionless
damping. The cutoff of fast modes corresponds to the scale 
where \( \tau _{fk}\gamma _{d}\simeq 1 \) and this defines 
the cutoff scale \(k_c^{-1} \). As we see from Eq.(\ref{lan}), 
the damping increases with \(\theta\) unless $\theta$ is close to $\pi/2$. 

Using the tensors given in Eq.(4) we obtain the corresponding \(D_{\mu\mu}\) for the CRs interacting with fast modes by integrating Eq.(\ref{genmu}) from \( k_{min} \) to \( k_{c} \) (see Fig.(\ref{fig:incom}b)). When \( k_c^{-1} \) is less than \( r_L \), the results of integration for damped and undamped turbulence coincides. Since the \( k_{c} \) decreases with \( \beta  \), the scattering frequency decreases with \( \beta  \).

Adopting the tensors given in Eq.(5), it is possible to calculate the scattering frequency of CRs in high \(\beta \) medium. For instance, for density \( n=0.5 \)cm\( ^{-3}, \) temperature \( T=8000\)K, magnetic
field \( B_{0}=1\mu  \)G, the mean free path is smaller than the resonant wavelength for the particles with energy larger than \(0.1GeV\), therefore collisional damping rather than Landau damping should be taken into account. Nevertheless, our results show that the fast modes still dominate the CRs' scattering in spite of the viscous damping. 

Apart from the gyroresonance, fast modes potentially can scatter CRs by
transit-time damping (TTD) (\cite{Sch98}). TTD happens due to the resonant interaction with parallel magnetic mirror force
\( -(mv_{\perp }^{2}/2B)\nabla _{\parallel }\mathbf{B} \). For small
amplitude waves, particles should be in phase with the wave so as to
have a secular interaction with wave. This gives the Cherenkov resonant
condition \( \omega -k_{\parallel }v_{\parallel }\sim 0 \), corresponding
to the \( n=0 \) term in Eq.(\ref{genmu}). From the condition, we
see that the contribution is mostly from nearly perpendicular propagating
waves (\(\cos\theta\sim 0\)). According to Eq.(4),we see that the corresponding correlation tensor for the magnetic fluctuations \(M_{ij}\) are very small, so the contribution from TTD to scattering is not important.

Self-confinement due to the streaming instability has been discussed by different authors(\cite{Ces,Chana,Long}) as an effective alternative to scatter CRs and essential for CR acceleration by shocks. However, we will discuss in our next paper that in the presence of the turbulence the streaming instability will be partially suppressed owing to the nonlinear interaction with the background turbulence.

Thus the gyroresonance with the fast modes is the principle mechanism for scattering cosmic rays. This demands a substantial revision of cosmic ray acceleration/propagation theories, and many related problems may need to be revisited. For instance, our results may be relevant to the problems of the
Boron to Carbon abundances ratio. 
We shall discuss the implications of the new emerging picture elsewhere.

\section{SUMMARY}

In the paper above we have shown that

1. Scattering by fast modes is the dominant scattering process provided that turbulent energy is injected at large scales.

2. Gyroresonance is the most important for pitch angle scattering. Transit-time damping (TTD) of the resonant waves is subdominant because the corresponding magnetic fluctuations are nearly perpendicular to the mean magnetic field.

3. The scattering frequency by fast modes depends on collisionless
damping for viscous damping, therefore it varies with plasma \( \beta  \).

\begin{acknowledgments}
We acknowledge valuable discussions with Benjamin Chandran, Jungyeon Cho, Peter Goldreich, Randy Jokipii, Vladimir Mirnov, Reinhard Schlickeiser, James Cornell, Simon Swordy. We thank  Joe Cassinelli, Elisabete Dal Pino, Vladimir Zirakashvili for useful comments. This work is supported by the NSF grant AST0125544.

\end{acknowledgments}


\begin{thebibliography}{10}
\bibitem{tensor}Apparently \( M_{ij}, C_{ij}\) are 3D matrixes. However, the third dimension is not needed for our calculations.
\bibitem{diffSch}These tensors are different from that in \cite{Sch98}.
The fact that the fluctuations \( \delta \mathbf{B} \) in fast modes
are in the \( \mathbf{k} \)-\( \mathbf{B} \) plane
place another constrain on the tensor so that the term \( \delta _{ij} \)
doesn't exist.
\bibitem{Dup}For \(D_{\mu p}\), the expression is only true for Alfv\'en modes. There are additional compressional terms for compressable modes.
\bibitem{static}Cosmic rays have such high velocities that the slow variation of the
magnetic field with time can be neglected.
\bibitem{Chanres}We counted only the resonant term in \cite{Chana}. The nonresonant term is spurious as noted by \cite{Chana}.
\bibitem{damp}Since the turbulence injection scale is less than the mean free path, only collisionless effects are involved.
\end{thebibliography}
\end{document}